# The solar fast dynamo action extended to A, Ap and Am type stars

## D. V. Sarafopoulos

*Department of Electrical and Computer Engineering,*

*Democritus University of Thrace, Xanthi, Greece*

sarafo@ee.duth.gr

2019

# Abstract

Initially, we present our original fast dynamo mechanism being potentially at work concerning the Sun and the solar type stars. Then, based on this new prototype model, a dynamo action model explaining the generation of the magnetic fields for A, Ap and Am type stars will be derived. We argue that their magnetic behavior is essentially due to locally exotic plasma properties **leading to large scale charge separation in their interior**. Their powerful magnetic fields are generated by the revolution speed of these excess positive or negative charges tightly associated with the formation of one or two so-called **Torus** structures. **The lifespan of the Torus' charge defines the star's activity cycle**. In the Sun's case, two Torus structures are formed preserving their charge for ~11 years. The Torus charging results from two processes: First, the star's differential rotation with its shear layers and second, the local plasma property that the plasma rotational speed ($u_e$) is equal or greater than the local light speed ($c′$) in the medium. If $u_e > c′$, then two nearby pocket charges (initially formed by the star's differential rotation) will be mutually attracted. That is, the attractive magnetic force ($q\mathbf{u_e}x\mathbf{B}$) will be greater than the repulsive electric one ($q\mathbf{E}$). Eventually, more and more charges gather within the Torus forced by its own powerful poloidal magnetic field. In addition, the same poloidal field, in the sub-photosphere shear layer, produces an excess charge (and therefore an azimuthally flowing current) generating the starspot activity. The solar birthplace of Torus, within the tachocline, is positioned at ~±45º latitudes; while the terminal latitudes are ~±10º. **Each Torus drifts equatorward increasing steadily its own rotational speed and, consequently, its own strength of poloidal magnetic field.** It is worth noticing that the excess charge rotation speed of Torus is always different from the rotation speed of the subsurface layer excess charge, and respectively the same with the value of the photosphere plasma, when both have the same latitude. **And the subsurface current steadily enhances with the Torus' drift**, given that its excess charge is subject to the gradually stronger poloidal field of Torus. In a solar type star but with antisolar differential rotation (i.e., the starspots move to higher latitudes), the tori are probably formed at ~±10º latitudes and then drift all the way to ~±45º. A single Torus, over the equatorial plane, is probably formed in the interior of an Ap type star producing its strong and global dipole-like magnetic field. An intense radial gradient of plasma density appropriately located can play the same role with the radial differential rotation.

# 1. Introduction

We aim at introducing the basic adopted concepts building up a convincing mechanism generating the prominent magnetic features observed in Ap, Am and A stars. In our view, the heart of the dynamo action for all these stars is the so-called "Torus" structure with its own exotic properties, as first introduced by Sarafopoulos (2017) for the case of the Sun. A homopolar generator in the star's interior may be probably developed, but it does not certainly produce any fast dynamo action. A unique structure with novel properties and radically new behaviour is needed; and for this reason we suggest the Torus formed in the star's interior and carrying a tremendous amount of net positive or negative charge. In this approach, _**a charge separation process is the cornerstone of the model**_. We remind that, in the case of the Sun, not a single but two toroidal structures have been proposed directly related to the two sunspot zones symmetrically situated with respect to the equatorial plane. Furthermore, the charge (inhomogeneously distributed along the Torus) is subject to the rotational speed of the star and produces a strong toroidal current and an exceptionally intense poloidal magnetic field affecting the Torus nearby space. The basic mechanism related to the Torus' formation remains, in principle, the same as in our preceded work; however, we introduce here new fundamental elements and modify a few ideas of secondary importance. **We also remove a few common misunderstandings** related to the behaviour of the Torus entity immersed in a plasma regime subject to differential rotation. It is emphasized that the starspots (much like the sunspots) are not directly produced by the toroidal-azimuthal current of Torus and its own associated poloidal magnetic field; in contrast, the spots are directly related to a subsurface charge layer generated by the combined action of the Torus magnetic field and the local rotational speeds with the appropriate gradients. Rotation speed shear layers are deduced through the graph of rotation rate as a function of fractional solar radius, at selected latitudes. Actually we use the solar diagrams ($\Omega/2\pi$, $r/R_\odot$) resultant from helioseismology (Kosovichev et al., 1997; Schou et al., 1998) and existing in solar textbooks. If the Torus is charged positively, then the subsurface layer is charged negatively and vice versa.

Magnetic field generation by differential rotation is usually regarded as a process operating in the convective zones of stars. A "convective stellar dynamo" (e.g. Parker 1979) is then a field amplification process in which differential rotation stretches field lines into a toroidal field ($\Omega$-effect). Convective fluid displacements create a new poloidal field component ($\alpha$-effect) by putting kinks into the toroidal field lines. The stretching of these kinks produces a new toroidal component amplifying the existing toroidal field and thus "closing the dynamo loop". However, concepts significantly deviating from this mainstream approach are already expressed. For instance, one can read that "the solar cycle, generally considered as the classical case of a convective dynamo process, **is probably not driven by convective turbulence at all**" (Spruit, 2001). Also in the same work, it was clearly argued that "the generation of a magnetic field in a star requires only one essential ingredient: a sufficiently powerful differential rotation. The recreation of poloidal field components, which is needed to close the dynamo loop, can be achieved by an instability in the toroidal field". In our approach the convection does not play any significant role; conversely, we emphasize the consequences of differential rotation. Furthermore, <u>*we absolutely distanced from any MHD treatment*</u>. In parallel, it is commonly underlined that the open questions are extremely challenging; for instance, it is well known that the A-type stars are fully radiative and thus are not expected to harbour a magnetic dynamo, and the questions of how A-type stars originally acquire their magnetic fields and why only 10% of stars exert a magnetic field still remain unanswered (e.g., Kochukhov, 2013; Ferrario et al., 2014). It is generally believed that a prescribed fluid velocity field would lead a small seed magnetic field to grow exponentially in time (Brandenburg and Subramanian, 2005; Miesch, 2005; Ruediger, 2008; Fan, 2009; Charbonneau, 2010; Jones et al., 2010; Cheung and Isobe, 2014). In a very highly conducting fluid, to a good approximation, the magnetic field strength is considered to grow in proportion to the stretching of the magnetic field lines. <u>***In our approach, in contrary, the fundamental physical entities are both the electric current and the magnetic field***</u>; a toroidal electric current easily could produce the needed poloidal magnetic field. The concepts of "stretching, twisting and folding magnetic field tubes", although attractive, prevent us from determining the involved electric currents. Our primary task is to detect the existing charge separation mechanisms, which inevitably will produce charge circulation paths and potentially develop poloidal magnetic fields. We are convinced that all the abstract and elegant tools, together with the best elaborated concepts based exclusively on the magnetic field, will lead to an endless chaotic perplexity. The Ω- and α-effects associated with a recycling process of the magnetic field will not finally succeed giving a thorough and convincing dynamo action solution. Voices asserting that "recent data have thrown into sharper relief the vexed question of the origin of stellar magnetic fields, which remains one of the main unanswered questions in astrophysics" (Ferrario, 2015) may demand a deeper revision on the mainstream approach methodology. Really, we alienate from every MHD-based solution; and we stress, again, that *it is imperative for us to define the major involved currents instead of defining prescribed fluid velocity fields*. After all, it is absolutely legitimate, within an academic impartial environment, to scrutinize every promising approach.

In astrophysics, chemically peculiar (CP) stars are stars with distinctly unusual metal abundances, at least in their surface layers. Ap stars are chemically peculiar stars of type A which show overabundances of some metals, such as strontium, chromium and europium. These stars have a much slower rotation speed than normal A-type stars, although some exhibit speeds up to about 120 km per second. Babcock (1960) discovered a huge 34.5 kG (i.e., 3.45 T) magnetic field, in the star HD215441, known since then as "Babcock's star". Despite success in the years that followed in increasing the number of known magnetic stars, HD215441 remains the record holder for the main sequence stars with the strongest magnetic field. The Ap star HD75049, has the second strongest magnetic field with a maximum value reaching 30 kG (Elkin et al., 2009). In most cases of Ap stars a magnetic field which is modelled as a simple dipole is a good approximation. For instance, variations of the mean longitudinal magnetic field for the HD75049 can be described to first order by a centred dipole model with an inclination i = $25^o$, an obliquity β = $60^o$ and a polar field $B_p$ = 42 kG (Elkin et al., 2009). HD75049 has rotation period 4.049 d, radius 1.7 $R_\odot$ and displays moderate overabundances of Si, Ti, Cr, Fe and large overabundances of rare earth elements. As usual in physics, analysis of extreme cases may provide key information; hence, there is great interest in the stars with the strongest magnetic fields to help solve many problems in Ap stars. Every SrCrEu and Si-rich Ap star has the field of at least 300 G (Aurière et al., 2007). Simultaneously, there exists another group of CP stars (i.e., Am on the cooler side and HgMn/PGa on the hotter side) for which no credible evidence of magnetic fields has ever been presented. Thus, there exists a magnetic dichotomy among intermediate-mass A and B stars: strongly magnetic objects share the H-R diagram with the stars deemed to be completely void of surface magnetic fields. Recent improvements in the observational techniques of stellar magnetometry supported this magnetic dichotomy paradigm, demonstrating that every SrCrEu and Si-rich Ap/Bp star has the field of at least 300 G. The latter is further discussed in the following paragraphs.

Stellar magnetism studies carried out during six decades since the discovery of global magnetic fields in peculiar A stars have firmly established the bimodal character of the incidence of magnetic fields among intermediate mass main sequence stars. Perhaps the best illustrations of the bimodal nature, of the rotational distributions of A stars, come from Abt and Morrell (1995) and Zorec and Royer (2012). From observations, it is then clear that at least two Maxwellians are required to describe the distribution empirically (Abt and Morrell, 1995, look at their figure 4), with the slowly rotating population being chemically peculiar, that is, all Am or Ap stars.

In summary, the following major conclusions emerge from the recent magnetic field studies of peculiar and normal A stars:

1. All Ap stars are magnetic, with a minimum dipolar strength of 300 G.
2. There is a "magnetic desert" between 300 and ~10 G for A stars. Below this range, Vega-like fields can exist in the majority of stars.
3. Observations of Vega (Lignières, et al. 2009; Petit et al., 2010) and Sirius (Petit et al., 2011) point to an entirely new manifestation of magnetism among A stars. These fields probably exist in all intermediate-mass stars and are weaker by about two orders of magnitude with respect to the 300 G lower limit of the Ap-star magnetic field. Vega shows a relatively complex surface field structure, dominated by a polar field concentration where the field reaches 3 G.

Two main scenarios are usually proposed to account for the origin of stellar magnetic fields. The first option attributes a fossil nature to the magnetic field, in the sense that the field is inherited from star formation or an early convective evolutionary phase, the strength of which has been amplified during stellar contraction (e.g. Moss 2001). This first model is generally preferred to account for the strong magnetism of CP stars, since it can be reconciled with the observed simplicity of their field geometries (dominated by a dipole). A second option invokes the continuous generation of the magnetic field through dynamo processes active either within the convective core (Brun et al. 2005) or in the radiative layers (e.g. Lignières et al. 1996, Spruit 2002).

The existing dichotomy between strong and ultra-weak magnetic fields among intermediate mass stars (Lignières et al. 2014) has prompted Braithwaite and Cantiello (2013) to argue that such fields could be the remnants of fields already present or formed during or immediately after the star formation stage. Hence, these fields would still be evolving on a timescale that is comparable to the age of the star. In this context, all intermediate and high mass stars, lacking strong fields, should display sub-gauss field strengths that would slowly decline over their main sequence lifetime. Certainly, an important clue to distinguish between the fossil or dynamo hypothesis would be to investigate the long-term stability of the observed magnetic geometry, as a dynamo-generated field is likely to experience some temporal variability (Petit et al., 2010).

Since we distance from the mainstream treatment of dynamo action using the MHD theory, a prologue "section 2" is following in which the two key components of our approach model are emphatically exhibited. Moreover, given that the solar dynamo action is always assumed as the prototype for all the star models, we briefly present our solar dynamo mechanism in "section 3" (as it is established in close cooperation with the solar differential rotation). Then, we describe the dynamo action potentially at work for a typical Ap star (in subsection 4.1), an Am star (in subsection 4.2) and an A star like Vega (in subsection 4.3); all these mechanisms are activated within the radiative envelope. In addition, all the generated magnetic fields are likely to experience some temporal variability or may carry signatures from an evolutionary process; issues open to future research and possible observational corroboration. Within the latter frame, we discuss our dynamo action model in section 5.

## 2. The heart of the proposed dynamo action model

In our view, the cooperation of two mechanisms essentially produces one or two "Ring Currents" in the star's interior; that is, one or two distinct entities with huge toroidal currents. **Each so-called "Torus" carries a large amount of net charge moving with the local rotational speed of the star**; hence, the resultant huge toroidal current forms an exceptionally intense and prevailing poloidal magnetic field that causes the exotic properties of the Torus. The latter has (not unjustifiably) led Sarafopoulos (2017) **to characterize the Torus as a "giant superconductor", having a periodic lifetime of approximately 11 years**. As a matter of fact, the Torus organizes the overall magnetic behaviour of the star, where the periodic reversal in magnetic field direction is a prominent feature. Bellow, we shall briefly exhibit in detail two natural mechanisms, which can be considered as responsible for this extraordinary magnetization effect.

The first mechanism is due to the differential rotation of the star. Certainly, one is more familiar with the solar case, which is always our prototype approaching the star's magnetic behaviour. For this reason, the equivalent circuit topology and all the features associated with the differential rotation of the Sun will be adopted in the following. In **Figure 1**, we show a radial solar slab, where the vertical Z-axis divides the space in two regions with different rotational speeds: The left-hand region (R1) with inward speed $u_{e1}$ and the right-hand one (R2) with inward speed $u_{e2}$. We assume $u_{e1}<u_{e2}$, while the passage from R1 to R2 mimics the passage from the tachocline to the convective zone (CZ). All the electrons, affected by the northward directed seed magnetic field ($B_{seed}$), move to the left. **A reaction which is similar with what takes place in a homopolar generator**, where the Lorentz force $\mathbf{F}_L = -e\mathbf{u}_e \times \mathbf{B}_{seed}$ finally generates an electron current; a current that can reach tremendous intensities. Moreover, **the reduction of electron speed due to the solar differential rotation causes a negative charge accumulation** in the layer adjacent to the transition surface from R1 to R2 (red-shaded area). Consequently, in this way a large amount of net negative charge rotates eastward and produces a giant poloidal magnetic field. In parallel, at that point the second mechanism, discussed in the next paragraph, plays its own catalytic role. The rate of charge accumulation is accelerated;

that is, the second mechanism acts as a charge amplifier. Finally, the two mechanisms cooperate producing much higher densities of the excess charge.

The second mechanism leads to a very peculiar behaviour: **Charges of the same sign are mutually attracted.** The latter is potentially accomplished if the magnetic force (**$F_m$=-e$u_e$xB**) is greater than the electrostatic one (**$F_e$=-eE**). It is well-known that two charges of the same sign, moving translationally, are mutually attracted (as elementary currents) and repulsed (as electrostatic charges). And the above condition (i.e., |$F_m$|>|$F_e$|) is satisfied if $u_e$>c′, where c′ is the local phase velocity for an electromagnetic disturbance. The reader can found this condition in every textbook of electrodynamics; it is also explained in the preceded work of Sarafopoulos (2017). Actually, in our case, **c′ is identical to the radial diffusion velocity of photons** in the actual medium of propagation and $u_e$ will be **the average rotational speed of the excess charge.** Apparently, the value of c′ is very small within the tachocline. The limit $u_e$=c′ is drawn with a vertical red-dashed line. Thus the accumulated electrons (as we previously have argued about with the supposed $B_{seed}$) are subject to the condition $u_e$>c′; therefore, **the abundant electrons are mutually attracted!** The volume charge density of electrons dramatically increases leading to the exotic magnetic behaviour of stars. In the case of the Sun, two distinct Torus structures are formed (within the tachocline) and symmetrically positioned in respect with the equatorial plane. The Torus may be positively charged if electrons are removed from its interior.

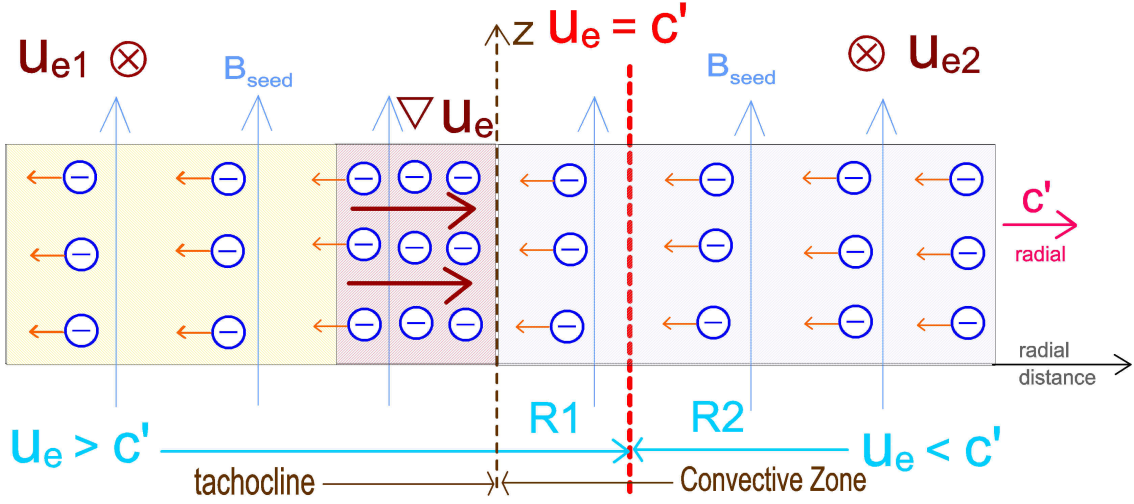

*Figure 1. A schematic showing a radial slab of solar plasma with a shear layer of rotational speed characterized by the gradient $\nabla u_e$ (red-shaded area). In this layer (formed adjacent to the line separating the tachocline from the CZ) electrons are accumulated, while moving inward. The rotational speed points inward and the seed magnetic field $B_{seed}$ northward. The red-dashed line defines the condition $u_e=c'$, where $c'$ is the radial speed of photons. Within the region of $\nabla u_e$ characterized by $u_e>c'$, the abundant electrons will be mutually attracted.*

The first suggested mechanism, in this subsection, is associated with what is happening in a conventional homopolar generator related to a **slow dynamo action**. In a laboratory workplace, one could obtain a similarly charged ring, if a disc composed of two different conductivities is appropriately constructed. However, **the fast dynamo action** is achieved only via the condition $u_e>c'$, dictating that charged particles with the same sign become mutually attractive. And the latter may remind us the coupling process producing "Cooper pairs" of electrons and the resonant behaviour achieved within a superconductor.

From these two mechanisms only the second one is sufficiently developed in the preceded work of Sarafopoulos. However, the differential rotation seems to play a key role. And many other significant aspects related to the differential rotation will become apparent in the following paragraphs.

## 3.1. Dynamo model for a solar type star

The purpose in this section is to put forward our approach concerning the dynamo action at work for a solar type star and, in this way, to establish a fundamental scenario which will become our prototype model for the normal and peculiar A-type stars, obviously after the appropriate modifications. The solar model exhibited in this work preserves the basic concepts introduced in our first approach (Sarafopoulos, 2017); however, it is differentiated in a few aspects. For instance, as it is stressed in the preceded subsection, the crucial role of the solar differential rotation is better recognized and integrated in this work. The presentation is brief and covers a complete 22-year solar-type cycle. The schematic **Figure 2** illustrates the basic topology of distinct plasma regimes and the distributions of currents and magnetic fields in a meridional cut-plane all the way from the RZ to the photosphere. In the first place, we point out three topological key features of crucial importance as follows:

1). We assume the existence of a seed magnetic field $B_{seed}$, being a necessary triggering condition for the solar-type cycle. $B_{seed}$ is taken northward directed and penetrates all the space outside the RZ. Certainly, after the first sunspot cycle, the seed field does not play any role in developing the dynamic model.

2). The vertical surface marked by dashed-red line is defined by the relation $u_e=c′$, where $u_e$ is the local (average) electron rotational speed and $c′$ is the local speed for an electromagnetic disturbance. When $u_e>c′$, the magnetic force is prevailing upon the electrostatic one, the reverse is happening when $u_e<c′$; as it was already emphasized. This surface intersects the equatorial plane at $r \approx 0.74$ $R_\odot$, given that the rotational speed shear layer is extended from 0.66 to 0.74 $R_\odot$; that is, the surface $u_e=c′$ is positioned in the outer layer of tachocline. The precise knowledge of the plasma differential flow within the solar envelope is an essential feature in our model.

3). **Each Torus (orange coloured solid circle) birthplace occurs at a latitude of ~45º**, just inside the line $u_e=c′$. The reason explaining this specific value of latitude is given later on. The observer, at present, is assumed being far away from the sun; consequently, we adopt a heliocentric system of reference.

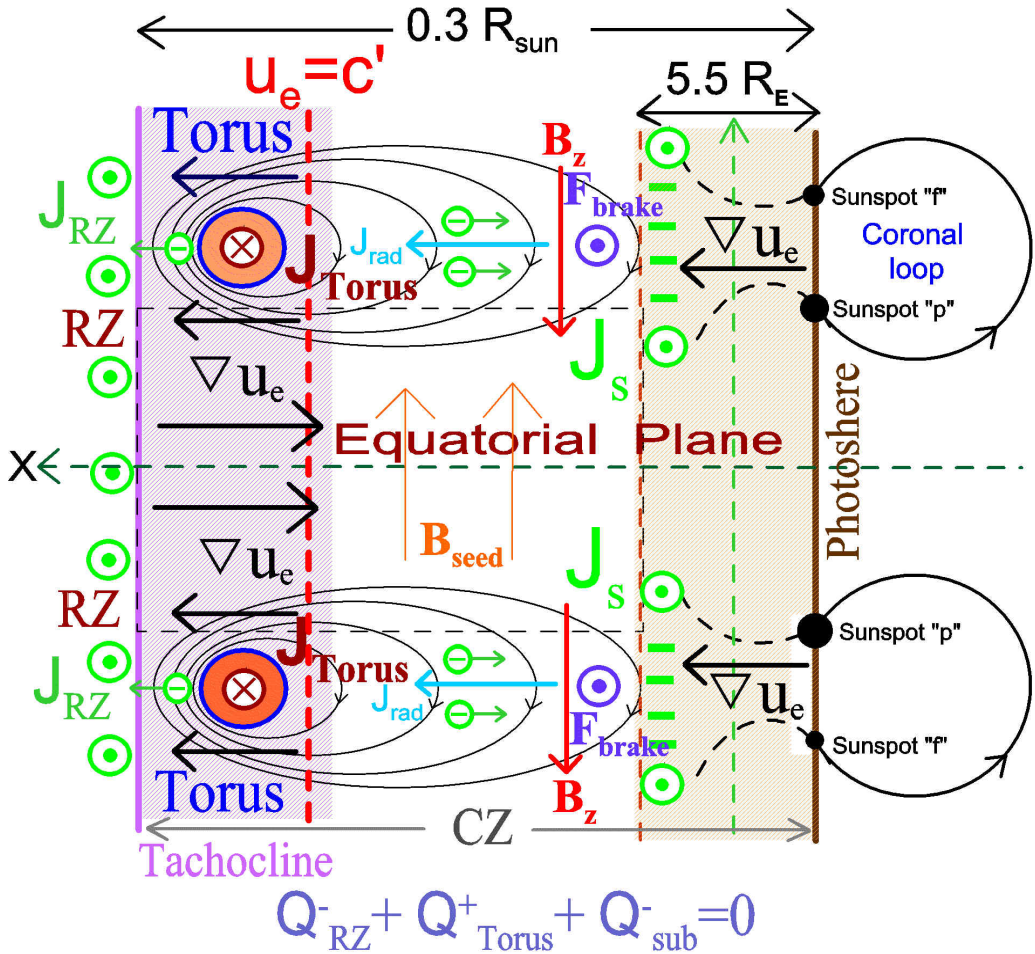

*Figure 2* *illustrating the solar type dynamo action with northward directed seed magnetic field ($B_{seed}$). The north and south toroidal structures (being positively charged here) carry eastward flowing currents ($I_{Torus}$), whereas on the surface of the RZ and in the photosphere subsurface layer the electron currents ($I_{RZ}$ and $I_S$, respectively) are westward directed. The overall charge from the two tori equals the surface charge of the RZ plus the photosphere subsurface charge. The azimuthally varying subsurface current creates the visible magnetic loops potentially emerging from the photosphere. The rotation speed gradients, resulting from the solar differential rotation, are shown within the tachocline and the subsurface layer as $\nabla u_e$. The surface defined by $u_e=c'$ is indicated with the dashed-red line. The braking force is $\boldsymbol{F}_{brake}=\boldsymbol{J}_{rad}\times\boldsymbol{B}_z$.*

The Lorentz force **F** = -e**u**$_e$x**B**$_{seed}$ steadily moves electrons inward. Obviously their collision frequency is much greater than the gyration frequency around the **B**$_{seed}$. Most importantly, at a site characterized by r ≈ 0.74 R$_☉$ and latitude of about ±45º, the electrons are subject to an intense acceleration. The reason is that, at that place, there is an inward directed strong gradient for the rotational speed; the rotational frequency (Ω/2π) changes from ~410 to ~430 nHz. The latter is apparent in **Figure 3**, reflecting our contemporary knowledge for the solar differential rotation effect (e.g., Kosovichev et al., 1997; Schou et al., 1998; also look at the figure 6.7 of Rüdiger and Kitchatinov, 2009). In this figure, the internal rotation in the Sun is sketched, showing differential rotation in the outer convective region and almost uniform rotation in the central radiative region. The transition between these regions is called the tachocline with very large shear as the rotation rate changes very rapidly. The reader has to pay particular attention to the (Ω/2π, r/R$_☉$) diagram, displaying the sharp Ω/2π variations for four contours corresponding to 0º, 15º, 30º and 45º latitudes. Consequently, just inside the curve u$_e$=c′, a region with net positive charge is formed and progressively evolved into an early (positively charged) Torus, which is indicated (in the figure) with an orange coloured solid circle. The charge itself, being subject to the local rotational speed, gives birth to an eastward directed current with its own poloidal magnetic field. Furthermore, the poloidal field, expelling additional electrons, progressively increases the Torus amount of excess charge; a positive feedback action. It should be stressed that the quantity of toroidal charge is azimuthally varying. Consequently, the Torus is associated with an azimuthally varying current and its own poloidal magnetic field. In effect, the Torus becomes a powerful engine expelling electrons with a gradually increasing rate; and, obviously, the poloidal magnetic field is much greater than the pre-existing seed field. Besides, at the Torus' latitudes and within the CZ, the electrons steadily move toward the photosphere. In this way, an extended channel is formed with electrons flowing all the way from the Torus to the solar subsurface layer. A layer in which the rotational speed is characterized by an inward directed gradient (inferred from the solar differential rotation diagram). Thus, an electron accumulation is locally produced due to the local electron deceleration. **Ultimately, this westward current of the subsurface layer could produce sunspot** **pairs on** the solar surface and huge magnetic loops emerging from the photosphere (as shown in **Figure 2**).

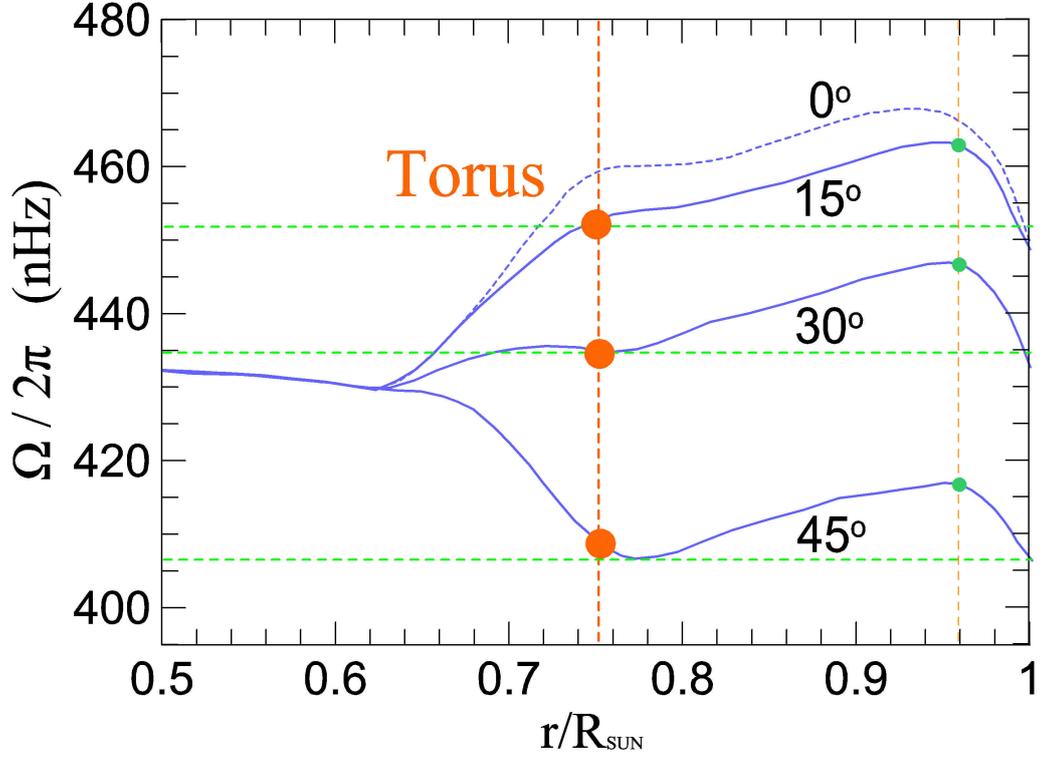

*Figure 3. Internal rotation rate in the Sun as a function of fractional solar radius, at selected latitudes. The differential rotation is apparent in the outer convective region, while we observe an almost uniform rotation in the central radiative region. The transition between these regions is called "tachocline". The orange coloured solid circle indicates the Torus shifting from the birth-place of 45º to 30º and 15º, in each 11-year cycle.*

If we pay attention to the north-south direction of the Torus neighbourhood, then the rate of removing electrons is greater from the equatorward side. Since the rotational speed monotonically increases toward lower latitudes (a response due to the solar differential rotation) and the Lorentz force is affected in a similar way, too. Thus, the Torus is basically shifting equatorward and the current density is always greater at its equatorward side. However, the drift motion halts at a latitude of about ±10º, as one could infer while inspecting the well-known solar "butterfly diagram". At that latitude, in the space between the two toroidal structures, the magnetic fields essentially eliminate each other due to their opposite orientations. In contrast, at radial distances greater than the two tori and close to the equatorial plane, the dominant magnetic field is southward directed; accordingly, none electron is allowed to move inward and reaching the RZ. Besides, in the region adjacent to the RZ the situation is radically different: In the space between RZ and each Torus, the (always northward directed) magnetic field abruptly increases. The magnetic field due to the toroidal current is superimposed over the field produced by the RZ current (due to an excess of negative charge). In effect, the surface charge of the RZ steadily increases its value and eventually **the Torus cross sectional area enhances**. That is, the Torus inner edge is gradually extended closer and closer to the RZ. Thus, at the latitude of about ±10º, each Torus turns stationary and begins merging with the RZ. A conductive tunnel is formed between the Torus and the RZ; while the northward directed magnetic field (preserving the two entities separately) is completely removed. Normally, between each Torus and the RZ a repulsive force is developed resulting from the oppositely flowing currents and the specific condition that the magnetic force is greater than the electrostatic one (in this region inward of the line $u_e=c'$). And one understands why both, **the Torus and the RZ could not simultaneously be negatively charged**. At that stage (with an open "conductive tunnel"), the RZ quickly will be switched to a new state carrying positive charge. The latter follows from the fact that the overall charge of the two tori is greater than that of the RZ. We plausibly assume that the positive charge from the two tori equals the negative charge of the photosphere subsurface layer plus the negative surface charge of the RZ. Therefore, at that point an 11-year sunspot cycle is completed and the magnetic field in the CZ is reversed to the south. In the new sunspot cycle each Torus will be negatively charged, while all the basic processes will follow a similar evolution, as in the preceded cycle. The newly created southward magnetic field (in the CZ) will move electrons toward the photosphere; in addition, within the tachocline and at ~45º latitude, these electrons will be subject to an intense deceleration. In this way, the resultant local electron accumulation will be the birthplace for the new (negatively charged) two Torus structures, very close to the $u_e=c'$ line. At the end of the new 11-year cycle the RZ will be negatively charged and a new 22-year cycle will begin.

A major new element in this work is that **the Torus begins be forming at ~45º latitude.** The equatorward drift of Torus is accompanied by additional crucial information, as we shall point out in the next subsection.

## 3.2. The same phenomenon evident by different observers

In this subsection, we scrutinize whether the same natural phenomenon preserves its identity for different observers. Usually there is a prompt reaction that the proposed model does not agree with the reality; in particular that a physical phenomenon (for instance an active sunspot pair) is visible only from a specific position. However, this is not the situation, if we really comprehend the whole scenario and pay attention to all the parameters contributing to the **sophisticated and well regulated system** being studied in this work. Finally, one will infer that two different observers suffice to demonstrate the functionality of the model for all the possible cases. The first observer is supposed to be positioned far away from the sun and the second one inside the Torus, while we focus on a typical sunspot pair (over the solar surface) activated by an existing subsurface current. Certainly, for the second observer the axes of the coordinate system are in parallel with those of the heliocentric one. The two observers must record exactly the same huge dipolar magnetic field structure supposedly related to the sunspot pair. We explain the situation in the next two subsections.

### *An observer far away from the sun*

It is underlined that the excess Torus charge, with its own revolution speed, builds up the powerful solar engine, producing all the sunspot activity in a cycle. From the preceded treatment, we infer that for a northward directed seed magnetic field, each Torus will be positively charged and the observer will record an eastward directed toroidal current. It has been already emphasized that the seed field, although having a triggering effect, however, the next solar cycles are not depending on it. In the space between Torus and photosphere, a southward directed poloidal magnetic field is developed due to the toroidal current. And this prevailing poloidal field, via the Lorentz force, expels electrons radially outward; in this way, a negative charge is finally raised in a photospheric subsurface layer. The reason for the latter is an abrupt drop of the angular rotational speed $\Omega$ within the subsurface layer, leading to an electron deceleration-accumulation, a process emphasized in subsection 3.1. The sharp drop of $\Omega$ is seen for all the four contours corresponding to 0º, 15º, 30º and 45º latitudes in the ($\Omega/2\pi$, r/R$_\odot$) diagram shown in **Figure 3**. The Torus is indicated with an orange coloured solid circle, while **the subsurface layer moves equatorward as tightly coupled with the Torus**. And this net-negative subsurface charge, observable as a westward current, probably produces the northward directed huge magnetic loop extended upward in the solar corona. During an 11-year cycle, as each Torus drifts equatorward, the same motion is done by the emerging short-lived sunspot pairs.

## *An observer positioned inside the Torus*

For an observer positioned inside the Torus, we know that **the Torus and the photosphere surface with the same latitude, ranging from 10º to 45º, have the same rotation speed $\Omega$.** The latter is apparent from the displayed ($\Omega/2\pi$, r/R$_\odot$) diagram; the dashed-green horizontal lines underline the above statement for three distinct latitudes. The supposed observer will be monitoring stationary charges within the Torus and, consequently, the Torus is disassociated from any toroidal current and poloidal magnetic field alike. Does it imply that the "powerful Torus machine" is completely inactivated? The answer is negative, as we shall exhibit below; however, additional key information is needed.

In the context of this work, the Torus' birth-latitude is positioned at ~45º and the terminal-latitude at ~10º, during the period of a solar cycle. And this equatorward drift of the Torus inaugurates an additional process worthy of attention: **The excess charge enclosed within the drifting Torus is subject to a monotonically increase of its rotational speed**, as it is inferred from the differential rotation diagram. The latter implies that the role of the equatorward drift is of paramount importance; the toroidal current, in effect, abruptly increases and the poloidal magnetic field alike. Furthermore, given that the poloidal magnetic field essentially determines the sign and the amount of the photosphere subsurface charge, then one infers that <u>**the subsurface current (and the solar activity as well) steadily intensifies with the Torus' drift**</u>. Finally, there is an extended stay of the Torus at ~10° latitude.

It is noted that the sign of the subsurface charge is determined by the poloidal magnetic field direction. Additionally, **the Torus plasma speed is always different from the subsurface layer plasma speed**. With positive charge in the Torus (resulting from a northward directed seed magnetic field) the subsurface layer will always carry negative charge always characterized by **higher revolution speeds**. The layer is formed around those $\Omega/2\pi$ values that are indicated by green-solid circles in the figure. Thus, an observer within the drifting Torus will always discern a subsurface westward current, from which all the sunspot activity is dependent. Additionally, as we shall see below there is one more critical parameter.

From the above reasoning one concludes that the Torus must achieve a mature state empowering it to produce and modulate the subsurface charge-layer. An observer positioned within the Torus, with an initial speed $u_e$ at the beginning of a new solar cycle, will measure a constant increase of the revolution speed as the time proceeds and the Torus drifts equatorward. As soon as the Torus achieves its mature state, the same observer will record an increment of speed, $\Delta u_e$, leading to an enlarged poloidal magnetic field. In conclusion, **an additional key parameter building up the poloidal magnetic field is the equatorward drift of the Torus itself**. Perhaps a few months suffice to give birth to a mature Torus with its associated subsurface current.

It is noted that the negative charge in the sub-photosphere layer has an increased rotation speed (i.e., $\Omega+\Delta\Omega$) as compared to the rotational speed $\Omega$ of the Torus. Thus, the net negative subsurface charge together with its eastward speed (for the supposed observer) will produce the same photospheric sunspot pair, as in the preceded case of observer. And **the sunspot activity will intensify as the sub-photosphere charge drifts equatorward increasing its speed** (i.e., a spatiotemporal variation is evident). In the next sunspot cycle, the whole scenario will be appropriately modified, but the final outcome will be the same. Obviously the role of the Torus drift, as emphasized in this subsection, is equally important for all the observers, and not only for those positioned inside the Torus.

## *An observer positioned on the photosphere surface*

This case is absolutely similar with the preceded one, since an observer positioned inside the Torus has essentially the same Ω with another positioned on the photosphere surface, given that both of them have the same latitude. And if one places an observer within the subsurface layer, then the positive Torus charge and a fraction from the negative subsurface charge moving backward will produce, in effect, the same magnetic loop.

## 4.1. Dynamo action model for an Ap star

It is underlined that a percentage of about 10% of all the A stars are Ap-type. Moreover, we assume that on the existing Ap population two separate mechanisms would be at work. The first mechanism is based on a differential rotation scheme similar to that observed in the Sun, while the second one is based on an intense radial density gradient. Both of the required conditions are fulfilled hard in A stars; and this is probably the reason that the Ap stars are only a small sub-group of A stars. We exhibit first the mechanism that resembles more with the solar case and then the second possibility, which may be the prevailing one. In any case, there is only a distinct main difference between the two mechanisms.

## *Dynamo based on supposed differential rotation*

The dynamo action for an Ap star has many aspects similar to those involved in the solar-type star treatment. The "Torus structure" remains the powerful engine; that is, the heart of the fast dynamo. However, it is not meaningless to briefly mention all the spatial and temporal elements used to build up a particular mechanism concerning the Ap stars. The seed magnetic field is considered northward directed. The four distinct types of surfaces (actually layers), which are depicted in **Figure 4**, are the following:

<u>First</u>, the stellar photosphere, being the outer limit of the radiative envelope: It is designated as surface **S1** and corresponds to the coolest, visible and completely ionized layer of all the spectrally classified A stars.

<u>Second</u>, the spheroid surface **S2** dictated by the condition $u_e = c'$. S2 is an ellipsoid by revolution obtained by rotating an ellipse about its rotational axis. The local rotational speed is indicated by $u_e$, while $c'$ is the local velocity for an electromagnetic disturbance (i.e., the local phase velocity of the light).

<u>Third</u>, the surface **S3** on which the speed of the radially inward (or outward) moving electrons (affected by the Lorentz force) becomes negligible, while the plasma density abruptly increases. The inward (outward) flowing electrons probably lead to the formation of a thin and distinct layer characterized by electron abundance (depletion), in the star's interior. This way a negatively (positively) charged shell is formed and marked as surface S3.

<u>Fourth</u>, the layer **S4** surrounding a huge "toroidal plasma cavity" with increased rotational speeds due to the supposed stellar differential rotation. This layer is characterized (at low latitudes) by an **inward (outward) directed gradient of rotational speeds** close to the star surface (at the inner edge of the CZ). Obviously we have adopted a solar type ($\Omega/2\pi$, $r/R_\odot$) diagram shown at the lower part of the figure for latitudes from $0°$ to $\pm 15°$. On the basis of these speed gradients, the whole shear layer is divided (in our sketch) into the red **S4a**- and green **S4b**-shaded areas.

.

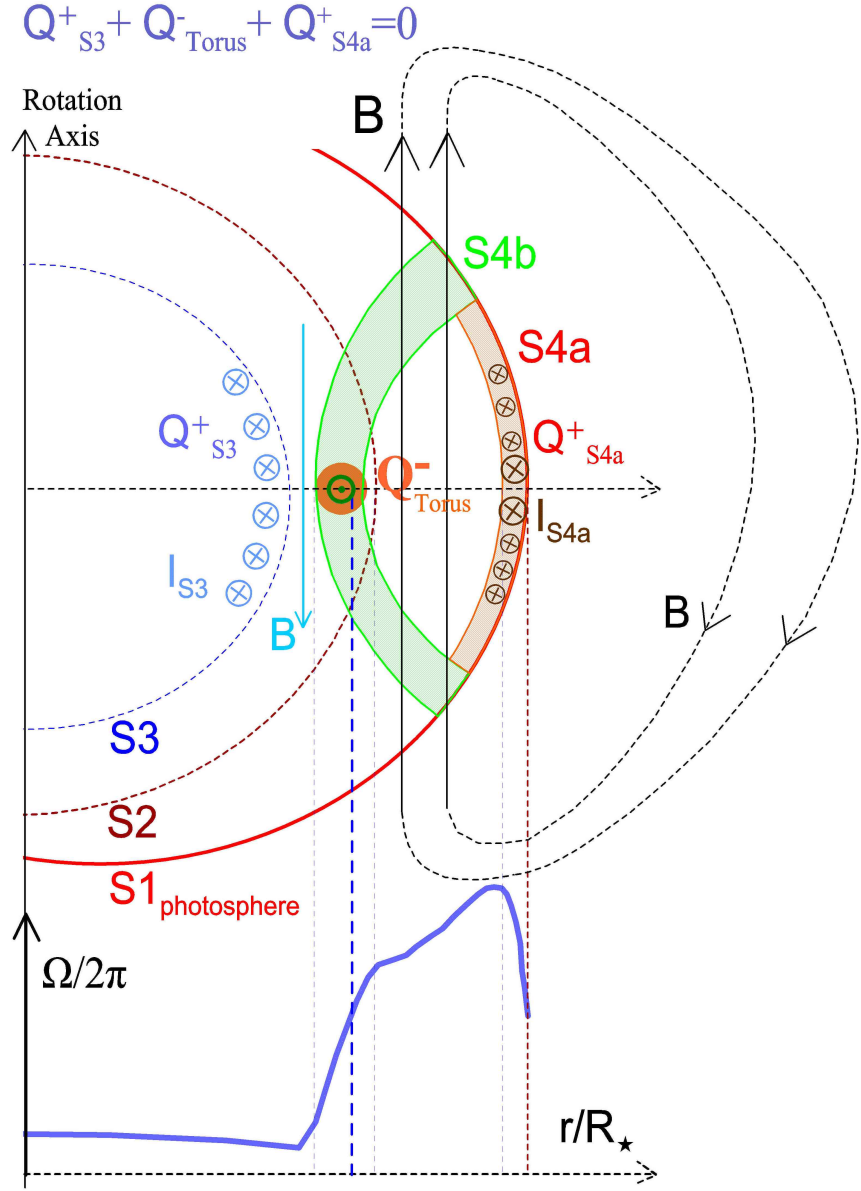

*Figure 4.* *Sketch corresponding to an Ap star with its own distinct layers. The single Torus (orange coloured solid circle) is formed (a) over the equatorial plane, (b) inward of the curve S2 defined by $u_e=c'$ and (c) within the rotational speed shear layer S4b. The Torus charge equals the overall surface charge of the surfaces S3 and S4a; i.e., $Q^-_{Torus}+Q^+_{S4a}+Q^+_{s3}=0$. The rotational speed variation at latitudes 0-**15°**, for an indicatively adopted solar type ($\Omega/2\pi$, $r/R_\odot$) diagram, is included in the lower part (blue-thick line). The dipole-like configuration of the magnetic field is represented by just two field lines.*

Furthermore, a few fundamental elements (being at the same time major key assumptions) for the Ap dynamo action, are the following:

a). The star is characterized by differential rotation; we adopt the solar radial profile of differential rotation to standardise the case. However, while the differentially rotating region for the Sun has a thickness of ~0.3 $R_{\odot}$, we consider that for an Ap star the corresponding thickness is significantly smaller. That is, the differentially rotating region is narrower in Ap stars. Although the Ap-type stars are characterized as slow rotators; however, their rotational speeds are much higher than those of the Sun.

b). ***Just one Torus structure is developed in the star's interior*** and over the equatorial plane producing an essentially dipolar magnetic field. Moreover, ***the Torus is displaced closer to the surface***.

c). The Torus, being the powerful dynamo engine, is formed within the layer of intense shear rotational speed designated as S4b.

d). Moreover, the Torus is formed at the intersection of the spheroid surface S2 (defined by $u_e=c′$) and the layer termed S4b.

e). The negatively charged Torus builds up an oppositely charged subsurface layer, the $Q^+_{S4a}$. In this sub-photosphere layer, the rotational speeds increase with the distance from the surface.

f). There is a northward directed seed magnetic field, which may dictate the sites of ion spots of the chemically peculiar Ap star.

After these lengthy introductory comments, we describe the dynamo action as follows: Initially electrons move inward motivated by the Lorentz force $\mathbf{F} = -e\mathbf{u}_e \times \mathbf{B}_{seed}$, in a similar manner as in the solar case. However, their deceleration within the layer S4b results in a local electron accumulation, producing eventually the negatively charged Torus ($Q^-_{Torus}$). The drop of the local rotational speed is apparent at the adopted ($\Omega/2\pi$, $r/R_{\odot}$) diagram shown in the lower part of **Figure 4**. Apparently, the $Q^-_{Torus}$ associated with a toroidal current and poloidal magnetic field steadily increases. The latter results since the attractive magnetic forces (due to the Torus current) overcome the repulsive electrostatic ones. The Torus is formed inward of the surface $u_e=c′$ and its own poloidal magnetic field becomes the dominant one, attracting more and more electrons inward. At the same time, the moving electrons leave a positively charged layer close to the star's surface, the $Q^+_{S4a}$. More specifically, the subsurface charge is mainly formed at the inner portion of the layer S4a (on the basis of the local $\nabla u_e$) and produces an intense eastward directed current.

The space between the negatively charged Torus and the positively charged subsurface layer is characterized by an intense northward directed magnetic field being the established dipolar field for an Ap star. The question is whether indeed this field is an unvaried, stable and long-lived structure. On this issue, *we assume that a periodic fluctuation of the magnetic field, ranging for instance from ~300 G to ~30 kG, may be at work.* Although the Torus is prevented increasing its cross-sectional area outside the surface S2, however, the Torus could be readily extended inward. The Torus pumps electrons from the layer S3 (which is positively charged by $Q^+_{s3}$) and steadily increases its cross-sectional area approaching the surface S3. The corridor between the Torus and S3 becomes gradually narrower, while the sandwiched magnetic field monotonically increases. As a consequence, this positive feedback unavoidably leads to the Torus collapse: Finally, the inward inflated Torus encounters the surface S3 and the achieved charge separation is destroyed in a large degree. An "electric discharge" takes place; however, the surface S3 will continue to accommodate a small amount of negative charge that equals the positive charge of the subsurface layer $Q^+_{S4a}$. It is plausibly assumed that $Q^-_{Torus} + Q^+_{S4a} + Q^+_{s3} = 0$. In this extreme state of minimum charging the dynamo is assumed able producing only ~300 G. Certainly, a new cycle of Torus charging will be immediately initiated and the magnetic field will gradually increase up to the upper limit of ~30 kG. In this scenario, we do not propose a solar type reverse of the magnetic field; instead, a periodic fluctuation of the magnetic field strength might be a new fingerprint of model. And each single cycle may be highly asymmetric; that is, the stars may be not evenly distributed in the range from 300 G to 30 kG. After all, the proposed variability is just a possible evolutional process; the periodicity may be a few hundred years, however, it is out of the scope of this work to attempt estimating a repetition time.

The positive (subsurface layer) charge density for an Ap-type star is greater than the solar subsurface layer density, since the Ap is completely ionized. In addition, there is a mechanism largely modulating the dipole-like structure of the magnetic field: We have already assumed that electrons are steadily pumped radially inward from the subsurface layer S4a and mainly from the lower latitudes. Thus, new electrons have to flow in from the northern and southern edges (of the layer S4a) replenishing the leaving population. In this perspective, ***the positively charged surface layer is significantly extended to higher latitudes. Consequently, a more dipole-like magnetic field configuration has to be expected.***

## Dynamo based on an intense radial gradient of density

It is extensively emphasized that an abrupt change of the radial rotation rate (occurred inward of the curve $u_e=c'$) would lead to the Torus formation. However, it should be underlined that a shear layer of speed is not the only possibility building up the Torus. An intense radial gradient of plasma density ($\nabla n$), inward of the curve $u_e=c'$, can play the same role leading to large scale charge separation and accumulation.

Obviously the Lorentz force motivates and displaces much more electrons out of the higher density region. Consequently, an existing negative or positive charge formed inward of the curve $u_e=c'$, may generate a Torus structure. With the supposed northward directed seed magnetic field and an inward $\nabla n$, the single Torus will carry positive charge, while the surface charge will be negative. In this magnetic field configuration an ion spot would be anticipated around the polar region or in a polar ring-type area. ***Eventually, the role of the density gradient may be equally important to the radial differential rotation***. The magnetic field will be a dipole-like one. The Torus latitude may be positioned somewhere between $0^o$ and $\pm 15^o$. Moreover, it should be stressed that ***the "Torus" may be azimuthally restricted*** in a region characterized by an intense radial density gradient: Even a local compression/expansion triggered by an impact event (i.e., a collision between astronomical objects) may suffice generating "locally" an extremely high value of toroidal current. In the latter case the magnetic field dipole axis will be oblique with respect to the stellar rotational axis. ***A further comment on the role of $\nabla n$ and the validity of the relation $u_e=c'$ is given in the subsection 5.8.***

## 4.2. Dynamo action model for an Am star

A prominent case of Am star is the Sirius A belonging to a binary star system and consisting of a white main-sequence star of spectral type A0 or A1 and a faint white dwarf companion called Sirius B. We suggest that such an Am star may be the evolutionary result from an Ap star; that is, an Ap star may terminate as an Am star. The latter although preserves (in a degree) its own chemically peculiar character, however, it stops being "a magnetic star". It belongs to a star category characterized by a very weak magnetic field of ~1 G. Which are the underlying principles for this switching process?

Unambiguously, every magnetic Ap star is subject to an intense and persistent magnetic braking process due to its own high magnetic moment. In our Ap model, the braking force **JxB** is generated from the radially outward flowing current and the northward directed magnetic field mainly in the space between the Torus and the sub-photosphere layer. And this drag force leads to a gradually lower rotational velocity. Consequently, we suggest that the surface S4b probably moves rapidly inward and the star finally flips to a new structure with two distinct toroidal structures; an arrangement adopted in the solar dynamo case. Moreover, the Torus structures are displaced deeper, while the rotational speed decreases. In the new topology, one can conjecture that the Am star could be subject to a solar-type periodic variation.

The above scenario may provide an answer to a longstanding question: The percentage of Ap/A stars in binary systems is 2%, whereas for single stars the same ratio is ~10% (e.g., Ferrario et al., 2015). The latter may occur due to the additional tidal braking force in binary systems. Thus, the Ap stars in binary systems are subject to an intense total braking force becoming very slow rotators. And the anticipated observable percentage of Am stars in binary systems has to be ~8%. Therefore, ***our suggestion gives an explanation to the observational evidence that Am stars are mainly observed in binary systems***. The Am star called Sirius A actually belongs to a binary system; and Sirius B may have increased the metallicity of its companion. The Am stars are very slow rotators and retain their peculiar character although in a smaller degree.

# 4.3. Dynamo model for the Vega class of weakly-magnetic stars

The repeated observational evidence that Vega possesses a weak photospheric magnetic field strongly suggests that a previously unknown type of magnetic stars exists in the intermediate-mass domain. Vega may well be the first confirmed member of a much larger, as yet unexplored, class of weakly-magnetic stars now investigable with the current generation of stellar spectropolarimeters (Lignières et al., 2009; Petit et al., 2010). Periodic variations of the signal are consistent with a rotation period of 0.732 days. Once averaged over the visible part of the stellar surface, the field strength does not exceed 1 G. This is comparable to the mean solar magnetic field. The average line of sight component of this field has a strength of −0.6 ± 0.3 G (Lignières et al., 2009), while the observed large-scale fields of all Ap stars appear to be larger than a few hundred gauss at the magnetic poles (Aurière et al. 2007). In 2015, bright starspots were detected on Vega's surface, the first such detection for a normal A-type star; and these features provide evidence for a rotational modulation with a period of 0.68 days (Böhm et al. 2015).

Vega has mass ~2.1 $M_\odot$ and its rotation velocity is ~236 km/s (Yoon et al., 2010) along the equator, which is 87.6% of the speed that would cause the star to start breaking up from centrifugal forces. This rapid rotation of Vega produces a pronounced equatorial bulge, so the radius of the equator is 19% larger than the polar one. The estimated polar radius is ~2.362 $R_\odot$, while the equatorial radius is 2.818 $R_\odot$ (Yoon et al., 2010), with $\Delta R=0.456\ R_\odot \approx 50\ R_E$. From the Earth, this bulge is being viewed from the direction of its pole, producing the overly large radius estimate. The pole of Vega (i.e., its axis of rotation) is inclined no more than five degrees from the line-of-sight to the Earth (Gulliver et al., 1994). The local gravitational acceleration at the poles is greater than at the equator, and the local luminosity is also higher at the poles. This is seen as a variation in effective temperature over the star: the polar temperature is near 10,000 K, while the equatorial temperature is 7,600 K (Peterson et al., 2006). This creates a thermal imbalance which is the cause of meridional circulation. The temperature gradient may also mean that Vega has a convection zone around the equator (Aufdenberg et al., 2006), while the remainder of the atmosphere is likely to be in almost pure radiative equilibrium.

Reiners and Royer (2004) and Reiners (2006) were able to detect differential rotation in three mid- to late-A stars. According to Balona (2013), using Kepler data, about 875 (40%) A-type stars show differential rotation similar to that of the Sun. Moreover, about 1.5 per cent of A-type stars are active, show flares and have starspots, although the sizes of starspots in A-type stars are similar to the largest sunspots. Rotational light modulation in Kepler photometry of K-A stars is used to estimate the absolute rotational shear. The rotation frequency spread in 2562 carefully selected stars with known rotation periods has been measured using time-frequency diagrams by Balona and Abedigamba (2016). They concluded that the time-frequency diagrams for A stars are not different from those in cool stars, further supporting the presence of spots in stars with radiative envelopes. One of the most important aspects of their study is the fact that rotational shear in A stars is clearly observed, although they stressed that the number of stars with no detectable rotational shear is largest among the A stars, comprising nearly 40% of the sample. This number is about 25% for F stars but drops to less than 5% for G and K stars. The same authors categorically underlined that "the long-held notion that spots should not exist in stars with radiative envelopes is clearly not correct. There are further indications that the current view of A star atmospheres is not correct". We shall not discuss the arguments expressed against these conclusions; it is out of the scope of this work. Future research will clearly cast light in these early results. However, we point out that these conclusions could be consistent with our dynamo scheme.

In the context of this work along with the observational evidence of Vega, one could infer that a dynamo action like that at work concerning the Sun may be a realistic scenario. Thus, **the Vega dynamo action could be essentially similar with the solar-type which has been described in subsection 3.1**. Two toroidal entities, one in each hemisphere, could potentially produce real starspots, while the mean magnetic field value is expected to be comparable with that of the Sun. In our approach, it is not of prime importance whether a convective zone is actually formed or not at Vega's lower latitudes; the convection process is not a prerequisite fundamental element directly coupled with the suggested dynamo action. The highly oblate geometry, probably causing an increased meridional plasma flow, implies that *the differentially rotating region is relatively thick*. Consequently, the two tori are probably formed far inward (from the photosphere); and this is the reason that the magnetic field is very weak. One may predict that starspot cycles could indeed take place in Vega, and the latter might be observationally established in the future.

# 5. Discussion

## 5.1. Dynamo action in Ap stars without convective envelope

We propose an alternative mechanism, among the existing ones, potentially producing the dynamo action in Ap stars. However, our approach is founded in a completely different framework; and it is meaningless to describe in detail the well-known hypotheses and discussions exhibited in review works and books, or to criticize the weak points in each hypothesis. Thus, we essentially bypass the discussion hypothesizing that the Ap magnetic field is either generated by convective processes in the core region of the star and then migrating through the RZ to the surface or that the magnetic field is based on fossils that were amplified as the star contracted out of the interstellar gas cloud.

**Our suggestion is essentially disassociated from every convection process**; and, therefore, it is radically distant from all the mainstream approaches related to the dynamo action models for stars. From this perspective, our first published work was focused exclusively on the Sun (Sarafopoulos, 2017); this work suggests a significantly improved version for the solar dynamo action extended to A-type stars. In the A stars, having radiative envelopes without convection, the usual alpha process for transforming toroidal to poloidal field would be absent and a solar type dynamo would not work. For instance, in the sentence expressed by Murphy (2015), **"without the surface convection, how can spotted-star signatures be generated in the light curve (of A stars)?"** essentially reflects the mainstream conviction. Thus, a physical mechanism for the production of magnetic field (and spots alike) in A stars that would particularly explain the observations like those of Balona (2013) **is lacking** (Murphy, 2015).

In our view, the strong poloidal magnetic field (being the heart of the dynamo action) is produced by an intense toroidal current flowing along the so-called "Torus" in the star's interior. We do not use the concepts of Ω- and α-effect or the buoyancy acting on a magnetic tube and every other notion or tool related to the MHD approach. We use exclusively the Lorentz force and the classic electrodynamics.

## 5.2. Ap dynamo action and white dwarfs

Fairly strong magnetic fields have been detected in white dwarfs. About 10% of white dwarfs have magnetic fields with strength in the range between about $10^5$ and $5 \times 10^8$ G (Landstreet et al., 2012). Currently, since a white dwarf is considered being a dense and degenerate electron gas with extremely high thermal conductivity (i.e., isothermal), convective motions needed for any dynamo action are improbable. This is the reason why the observed magnetic fields are considered as fossil fields: As the core of the progenitor star contracts and becomes degenerate, there will be an amplification of the field due to flux conservation.

In the context of this work, we could speculate that even in a white dwarf, the magnetic field would be produced by an existing so-called Torus structure, i.e., a huge toroidal current flowing in the star's atmosphere. A stellar evolution may also occur, but the magnetic field is not really the frozen-in flux from its progenitor star. As we have suggested in this work, an Ap star forms a single Torus in its outer layer of the radiative envelope; after the collapse process, its toroidal current may be well-preserved, although its cross-sectional area has certainly decreased by a factor $(R_{Ap}/R_{white\ dwarf})^2 \approx 10^4$. The radius of a white dwarf is of the order of 10 Mm, and if the magnetic field from a representative Ap star is supposed to be 1 kG, then the white dwarf-star would have a field of $\sim 10^7$ G.

## 5.3. A fresh look at the magnetic braking mechanism

It is assumed that magnetized stellar winds are important for extracting angular momentum from stars during the main sequence (Schatzman 1962) and likely during the pre-main sequence. In this perspective, for the case of the Sun, magnetic braking may result from solar wind material following the magnetic field lines that extend well beyond the stellar surface. This coupling exerts a torque on the surface layers of the Sun, and this slows down its rotation. Could such a process be active in massive stars possessing a sufficiently strong surface magnetic field? The latter is an open question on which several researchers are focused (e.g., Meynet et al., 2011). Given that our approach on the Ap dynamo action is radically different and we simultaneously use electric currents and magnetic fields, then (as expected) we introduce the well-known braking force **JXB** acting in the star's envelope layer.

More specifically, in the presented dynamo model, there is always a powerful poloidal magnetic field (due to the Torus current) and the radially flowing current particularly intensified around the Torus latitudes. Thus, the braking magnetic force, $F_{brake}$=**JXB** (shown in **Figure 2)**, will be always at work. Thus, in our view, **the braking force is the same that slows down the disc in a homopolar generator** or acts on the electro-mechanical brakes used to slow down or stop motion. For an Ap star, where the Torus is presumably formed close to the photosphere and strong magnetic fields are developed, the braking force will probably be more effective than in a solar-type star. And we discuss below the notion that the weakly magnetic CP stars, which are very slow rotators, may be evolved from Ap stars. Certainly, the braking force will heat up the star.

## 5.4. The suggested A-Ap-Am evolutionary scheme

***The concept is that the mechanism of magnetic braking forces a small fraction of A stars to become Ap-type magnetic stars; then, as the time proceeds, several Ap stars are transformed to weakly magnetic Am and HgMn stars***. The A stars are not related to any magnetic field or they possess a very weak average field. In contrast, according to our dynamo action model, an Ap star with its intense-global magnetic field is associated with just one Torus formed on the equatorial plane. Moreover, we assume that (besides the magnetic braking) an additional factor is acting in the above evolutionary scheme; and this is the stellar differential rotation observed in many stars. We assume that as the star slows down, the shear layer termed S4b (in subsection 4.1) is displaced inward; consequently, the overall magnetic behavior is essentially dictated from the intersection of curves $u_e$=c′ and S4b. In any case, the basic parameter is the time τ, representing the time of the hydrogen burning phase in the core of the star; the τ ranges from 0 to 1.Thus, three different geometries, dependent on the position of the layer S4b, are sketched in

33**Figure 5**. The curve $u_e=c'$ is marked with a dashed-red line. If there is no intersection between the two curves, a case corresponding to the green-shaded S4b layer with $\tau < 0.4$, then the A star does not form any Torus and completely lacks of magnetic field. If the common area of intersection is small, corresponding to the brown-shaded S4b layer with $0.4 < \tau < 0.8$, then a single Torus (marked by a brown solid circle) is formed and the star is an Ap-type one. Moreover, if the intersection area is large, corresponding to the red-shaded S4b layer with $\tau > 0.8$, then two toroidal structures (marked by two red solid circles) are probably formed in the star's interior and positioned symmetrically with respect to the equatorial plane; in this case, the star is a weakly magnetic CP star with solar-type dynamo action.

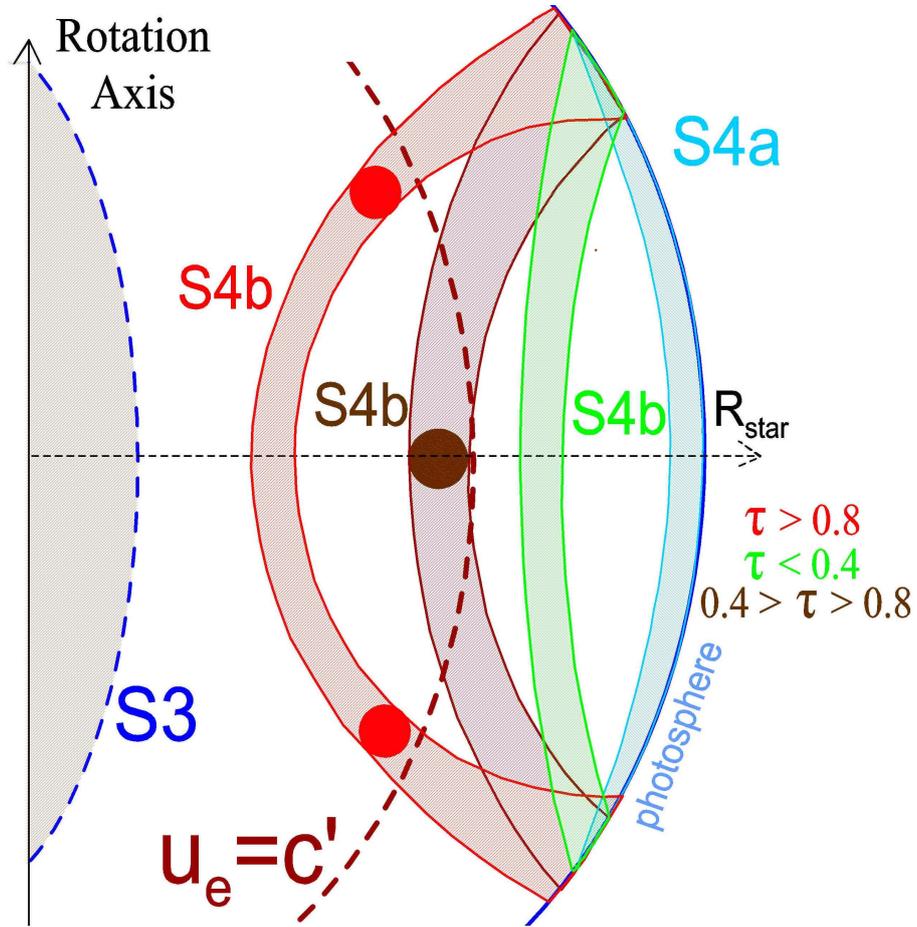

*Figure 5. The magnetic properties are largely dictated by the intersection or not of curves $u_e=c'$ and S4b. If there is no intersection (green-shaded area corresponding to $\tau < 0.4$), then the A star does not form any Torus and completely lacks of magnetic field. If the common area of intersection is small (brown-shaded area corresponding to $0.4 < \tau < 0.8$), then a single Torus (shown as a brown solid circle) is formed and the star belongs to the Ap category. Moreover, if the intersection area is large (red-shaded area corresponding to $\tau > 0.8$), then two toroidal structures (shown by two red solid circles) are probably formed in the star's interior and positioned symmetrically with respect to the equatorial plane; in this case, the star is a weakly magnetic CP star with solar-type dynamo action.*

## 5.5. Evolutionary process related to the work of Kochukhov and Bagnulo-2006

The work of Kochukhov and Bagnulo (2006) deals with the topic of evolutionary state of magnetic chemically peculiar (CP) stars on the base of a comprehensive statistical investigation of ~200 Ap-type stars. We would like to particularly discuss this work as being directly related to our approach. Their major results are almost perfectly in line with our thinking. A few key comments are the following:

1. In their figure 8b, there is a clear monotonic increase of the total magnetic flux $\Phi=<Bz>R^2$ and the star age $\tau$ (as defined in the preceded subsection 5.4), for all the three mass categories ($M \leq 2\ M_\odot$, $2M_\odot < M \leq 3\ M_\odot$ and $M > 3M_\odot$). In our model, the Ap star magnetic field strength fluctuates between maximum ($B_{mux}$) and minimum ($B_{min}$) values in a periodic fashion (that would correspond to several tens of years). Moreover, the $B_{mux}$ and $B_{min}$ values naturally increase with the age. The reason is that the sub-photosphere positive charge is preserved even at the time of Torus collapse; there is a charge cumulative effect. In the next cycle, an enhanced subsurface charge will increase both the $B_{min}$ and $B_{mux}$; and the same happens for all the successive cycles. Thus, while the magnetic field fluctuates from $B_{min}$ to $B_{max}$, the $B_{min}$ and $B_{max}$ values increase with the age $\tau$. Therefore, their figure 8b seems to comply with our scenario.

2. We infer that their statistical relation of $\Phi$ and $\tau$ means that the magnetic braking mechanism (as it is specifically defined in the preceded subsection 5.3) is at work in all the three star-mass categories. The latter is further corroborated from their figure 8c, where the star period (P) abruptly increases with the age ($\tau$), when $M<3M_\odot$; the latter is more profound for the $M<2\ M_\odot$ category. In contrast, the period very slightly increases with the age for $M>3M_\odot$. It seems that the magnetic braking is more efficient for stars with $M \leq 2M_\odot$; in contrast, the braking is inefficient for the massive stars with $M>3M_\odot$. That is, ***the whole star-life $\tau$ does not last long enough for producing very slow rotators in stars with mass greater than $3M_\odot$.***

3. From their figure 10 (right panel), it is very clear that slow rotators are exclusively produced from stars with M< $3M_\odot$. In our model, a fraction from these slow rotators would be readily converted to weakly magnetic Am and HgMn-type stars. And essentially none star from the category M>$3M_\odot$ should be anticipated to terminate as non-magnetic star. In their figure 8c, a very significant fraction of very slow rotators are produced for M ≤ $2M_\odot$. In the category $2M_\odot$ < M ≤ $3M_\odot$ very slow rotators are also observed, whereas in the third category M > $3M_\odot$ none star has become very slow rotator. The increase of the period P with the total flux Φ is also apparent in their figure 9 (middle panel).

4. From their figure 5 it appears that the higher mass stars (M>$3M_\odot$) are homogeneously distributed in fractional age. In the group of stars of intermediate mass, younger stars seem less numerous than older stars. Among stars with M ≤ 2 $M_\odot$ the shortage of young objects is more evident. In our scenario, the shortage of young stars (τ<0.4), in the category of M ≤ 2 $M_\odot$, is due to the needed time slowing down the rotational speed (via magnetic braking). In our view, a slow rotator Ap star, with its intense and well organized global magnetic field, evolves from an A star after reducing its angular momentum. The shortage of older stars (τ>0.8), with M ≤ $2M_\odot$, is attributed to the fact that a fraction from the very slow rotators are converted to CP non-magnetic stars.

5. How could we interpret the homogeneously distributed stars with mass greater than $3M_\odot$ versus their fractional age? In this category, we typically have low values of P and enhanced star radii. In effect, the rotational speed will always be relatively increased and the curve $u_e$=c′ is probably displaced radially outward. And in the new position this curve probably intersects the inner layer of shear rotational speed (i.e., the layer S4b) at very low latitudes. Therefore, a single Torus is probably built up.

## 5.6. The doubly-magnetic early-type close binary ϵ Lupi system

Shultz et al., (2015) reported the first detection of a magnetic field in both components of an early-type binary system (i.e., **the ϵ Lupi system**). The longitudinal magnetic field of the primary is about −200 G; that of the secondary about +100 G. Observations can be approximately reproduced by a model assuming the magnetic axes of the two stars are anti-aligned, and roughly parallel to their respective rotation axes. In the framework of this work, the fact of anti-parallel magnetic axes can readily be explained: If we suppose a dynamo action like that proposed for an Ap star, and an initially northward directed seed magnetic field, then the primary star will build up a southward directed magnetic field $B_{pri}$≈-200 G. Since the separation between the two stars is relatively small, we reasonably assume that the $B_{pri}$ will play the role of seed field for the secondary star producing finally the $B_{sec}$≈-100 G (with opposite polarity).

## 5.7. The antisolar-type differential rotation rate and the proposed dynamo action

The solar-like case of rotation rate decreasing with latitude is detected more frequently; whereas the anti-solar rotation is relatively seldom observed: Between eight observed anti-solar rotators, six belong to close binaries and two to giant stars with large thermal inhomogeneities on their surfaces (Strassmeier 2003).

Measurements and observations showing angular velocities increasing with latitude (and consequently the starspots moving poleward) can easily be interpreted on the basis of our proposed dynamo model. In the context of this work, for a star like the Sun but possessing an antisolar type differential rotation, a Torus could be generated at ~±10° latitude within the tachocline; then, it could drift all the way to the higher latitude of ~±45°. Such a poleward drift is dictated in this situation from the oppositely directed gradients of rotation speed. In **Figure 3**, you have to change the latitude of 45° with that of 15°, and vice versa. All the other processes remain unvaried as in the preceded treatment.

## 5.8. An additional qualitative look at the relation $u_e=c'$

We think that an additional comment focused on the relation $u_e=c'$ is useful. After all this relation constitutes the foundation of the suggested dynamo action. A qualitative approach, in spite of the fact that it gives an oversimplified picture of the real world, however, probably further clarifies our fundamental argument.

We assume almost completely ionized plasma characterized by density n and temperature T. In addition, the plasma is subject (Figure 6) to (a) a perpendicular to the page and inward directed bulk speed $u_e$, (b) an upward directed seed magnetic field **B**$_o$, (c) an intense $\nabla n$ perpendicular to **B**$_o$, and (d) a condition according to which the electron collision frequency is much higher than the electron gyrofrequencies. All these settings are probably satisfied in a specific radial distance in the interior of the star.

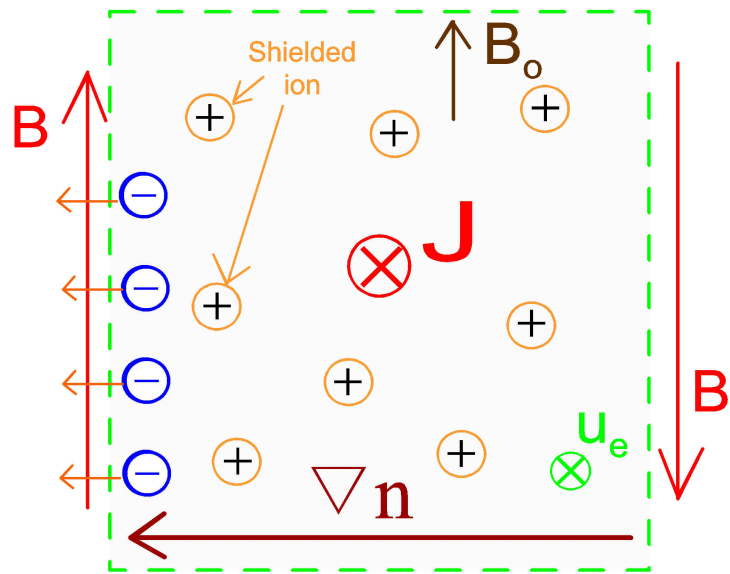

*Figure 6. $u_e$, $B_o$ and $\nabla n$ denote the bulk plasma velocity, the seed magnetic field and the intense gradient of the plasma density. There is a net outflow of electrons out of the shown volume (dashed-green line) motivated from the Lorentz force. The ions are shielded although a huge current J is established. Finally the electrostatic forces are minimized, whereas the attractive magnetic forces dominate. Under these circumstances the Torus is formed.*

The Lorentz force $\mathbf{F}= -e\mathbf{u}_e \times \mathbf{B}_o$ displaces more electrons from the side of higher plasma density. In the region of $\nabla n$ the ions remain stationary; in contrast, an electron depletion layer is developed. As anticipated the electrons respond collectively and each "positively charged center" is effectively screened outside of the Debye sphere. Consequently, ***each positively charged center in the plasma interacts collectively only with the charges that lie inside its Debye sphere***, its effect on the other charges being effectively negligible. Thus, the range of the electrostatic force is dramatically reduced and more electrons depart from the region of $\nabla n$. On the contrary, a new force comes into play. The gradual increase of the ion density produces a significant current density J flowing inward, since the bulk plasma velocity points inward. Most importantly, ***the unaffected poloidal magnetic field of the current J further facilitates the exodus process of electrons out of the region of $\nabla n$***. In addition, the high value of the plasma speed $u_e$ (e.g., being ~2000 kms$^{-1}$ for the solar tachocline) would produce huge current densities. Finally, the relation $u_e = c'$ means that the appropriate values of density and temperature probably satisfy the condition so that the attractive magnetic forces dominate against the repulsive electrostatic ones. Under these specific conditions the Torus structure is the splendid result.

## 6. Conclusions

We propose a promising fast dynamo action potentially at work on the Sun, and then extended to the A, Ap, and Am stars. Our model is completely distanced from the mainstream MHD approach. In parallel to a **slow dynamo action**, similar to that taking place in a conventional homopolar generator, we incorporate **a fast dynamo action** achieved via an exotic process taking place in the star's interior and building up a single or two toroidal structures carrying tremendous amounts of net charge. That is, **a large scale charge separation process is the cornerstone of the suggested model**. When $u_e > c′$, then distinct pocket charges, with the same sign, become mutually attracted and finally form the so-called **Torus** structure, where $u_e$ is the plasma rotational speed and $c′$ is the local light speed in the medium. That is, with this condition the magnetic attracting force ($q\mathbf{u_e} \times \mathbf{B}$) will be greater than the electric repelling one ($q\mathbf{E}$). Eventually, *the Torus reveals a behaviour like that encountered in superconductors*. The Torus fundamental property was first introduced by Sarafopoulos (2017); however, in this work the crucial role of the star's differential rotation is clearly recognized, incorporated and appropriately evaluated for the first time. Major findings for a solar-type dynamo model are the following:

1. The Torus birthplace is positioned at about ±45º latitudes, within the tachocline; while the terminal-latitudes are about ±10º. Each Torus drifts equatorward increasing steadily its own rotational speed and, consequently, its own powerful poloidal magnetic field.

2. The Torus' poloidal field, in the sub-photosphere shear layer (having the same latitude with the Torus) produces an excess charge (and therefore an azimuthally flowing current) generating the starspot activity.

3. The excess charge rotation speed of Torus is always different (the same) from the rotation speed of the subsurface layer excess charge (with the value of the photosphere plasma, when both of them have the same latitude).

4. The subsurface current steadily enhances with the Torus' equatorward drift, given that its excess charge rotational speed increases with the drift, too.

5. We speculate that in a star like the Sun, but showing an antisolar type differential rotation (i.e., the starspots move to higher latitudes), the Tori would be formed at $\sim\pm10^{\circ}$ latitudes and later they could drift all the way to $\sim\pm45^{\circ}$.

Although in the Sun's case, we have proposed two toroidal structures formed within the tachocline and directly related to the two sunspot zones symmetrically situated with respect to the equatorial plane, however, in the case of an Ap star a single Torus is probably built up producing its strong and global dipole-like magnetic field configuration. It is stressed that the key role of the differential rotation would be replaced by an appropriately located and inward pointing intense gradient of plasma density. Two major findings for the A-type stars are the following:

1. Given that in our treatment we use currents and magnetic fields at the same time, we can assume that the braking force of star is the same that slows down the disc in a homopolar generator. In our dynamo model, there is always a powerful poloidal magnetic field (due to the toroidal current) and a radially flowing current particularly intensified at the Torus latitude. Thus, the braking magnetic force, $\mathbf{F}_{brake}=\mathbf{J}\mathbf{X}\mathbf{B}$ will be always at work.

2. We propose a simple evolutionary scheme: The mechanism of magnetic braking forces a small fraction of A stars (being rapid rotators) to become Ap-type magnetic stars; then, as the time proceeds, several Ap stars are transformed to weakly magnetic Am and HgMn stars (being very slow rotators). That is, ***the dynamo-generated magnetic field is here recognized to be an important element affecting the normal stellar evolution.***

3. We speculate that even in a white dwarf the magnetic field would be produced by a charged Torus being a huge toroidal current flowing in the star's atmosphere.